\documentclass{aa}
\usepackage{graphicx}
\usepackage{lscape}
\usepackage{longtable}
\usepackage{natbib}

\newcommand{\gppr}{\stackrel{>}{\scriptstyle \sim}}
\newcommand{\lppr}{\stackrel{<}{\scriptstyle \sim}}

\begin{document}

   \title{A systematic study of X-ray variability in the ROSAT all-sky survey}

   \titlerunning{X-ray variability in the ROSAT all-sky survey}

   \author{B. Fuhrmeister
          \and
          J. H. M. M. Schmitt
          }

   \offprints{B. Fuhrmeister}

   \institute{Hamburger Sternwarte, University of Hamburg,
              Gojenbergsweg 112, D-21029 Hamburg\\
              \email{bfuhrmeister@hs.uni-hamburg.de}
             }


   \abstract{

We present a systematic search for variability among the
ROSAT All-Sky Survey (RASS) X-ray sources.
We generated lightcurves for about 30\,000 X-ray point
sources detected sufficiently high above background.
For our variability study different search algorithms were
developed in order to recognize flares, periods and trends, respectively.
The variable X-ray sources were optically identified with
counterparts in the  SIMBAD, the USNO-A2.0
and NED data bases, but a significant part of the X-ray sources
remains without cataloged optical counterparts.
Out of the 1207 sources classified as variable 767 (63.5 \%)
were identified with stars, 118 (9.8 \%) are of extragalactic origin,
10 (0.8 \%) are identified with other sources and
312 (25.8 \%) could not uniquely be identified with entries in optical catalogs.
We give a statistical analysis of the variable X-ray population and
present some outstanding examples of X-ray variability detected in the
ROSAT all-sky survey.  Most
prominent among these sources are white dwarfs, apparently single,
yet nevertheless showing periodic variability. Many flares from hitherto unrecognised
flare stars have been detected as well as long term variability in the BL Lac
1E1757.7+7034.

   \keywords{surveys --
             X-rays: general --
             stars: activity --
             stars: flares
               }
   }

   \maketitle
%

\section{Introduction}

The ROSAT X-ray observatory, launched in 1990,
carried out an all-sky survey during its first six months of operations.
This first X-ray imaging all-sky survey tremendously increased the number of
X-ray sources known at the time.  Variability is known to be one of the
key properties of the X-ray sky.
Almost all source classes, with the exception of supernova remnants,
clusters of galaxies and possibly white dwarfs, show variable X-ray emission.
Some source classes have already been searched for variability in the RASS data.
For example,  \citet{Haisch} studied the variability of RS CVn systems and other 
active giants detected in the RASS data,  
\citet{Greiner} searched for X-ray counterparts of
$\gamma$-ray bursts, and individual outstanding events have been reported.
\citet{Donley} searched systematically for galaxies with count rate
variability of more
than a factor of 20 by cross-correlating the RASS sources with data from
pointed observations and found five galaxies in this process. \citet{Grupe}
used a similar approach to search for variability in soft X-ray detected
active galactic nuclei. Furthermore
\citet{Stelzer1} investigated the X-ray properties of stars in the
Tucanae association using both RASS and pointed ROSAT data. They generated
lightcurves and found flares as well as irregular variability among young
stars. There are more examples for variability studies of individual sources
using the RASS data \citet{Schmitt1}, however, up to now no systematic search 
for variability in the total all-sky survey data has been carried out.

The first great advantage of the RASS data set is that it is
\textit{unbiased} in the sense that no specific objects or regions of the
sky were preferentially observed.  Therefore all types of objects can be 
analysed for variability and a statistical analysis of the variability 
properties can be performed in an unbiased fashion.  A second reason why
the RASS  data is very suitable for a variability search programs is the 
temporal sampling of the survey data. Due to the survey geometry all sources 
were scanned repeatedly for at least two days for up to about 30 seconds 
during each satellite orbit. Therefore the lapse time for each X-ray source 
is at least two days and much longer for sources close to the poles of the 
ecliptic.  Such a temporal sampling is almost never achieved in pointed 
observations.  Therefore for the study of variability on longer time scales 
there is a definite advantage of the survey data compared to  pointed 
observations, which are of course much deeper but typically sample 
variability on shorter time scales.

In this paper we will elaborate on the problem of finding variability
in the RASS data and identifying the variable X-ray sources with optical
counterparts. Sect. 2 of our paper describes the RASS data and the survey 
geometry, Sect. 3 presents the generation of the lightcurves and some 
problems arising in that
process; the different search algorithms used in our study
are explained as well. Sect. 4 discusses the
identification of the X-ray sources with optical counterparts via the SIMBAD and
USNO-A2.0 catalogs and the NED data base and
provides a statistical analysis of the variable
sources. Sect. 5 shows some examples of variability highlights we found
in this survey.

\section{The survey data}

The ROSAT satellite was operated in an almost circular low Earth orbit
at an altitude of 580\,km with an
inclination of $53^{\circ}$. It carried the \textit{Wide Field Camera}
for observations in the XUV and the X-ray telescope (XRT) for the measurement
of soft X-rays in the energy range of 0.1  - 2 keV,
corresponding to wavelengths of 120 - 6 \AA.  During the scanning phase
of the satellite the position sensitive proportional counter (PSPC) was
mounted in the focal plane of the XRT.   This is a photon counting
detector which registered the arrival time of
each X-ray photon, the position and the photon energy with quite
modest spectral resolution.

\begin{figure}
   \centering
   \includegraphics{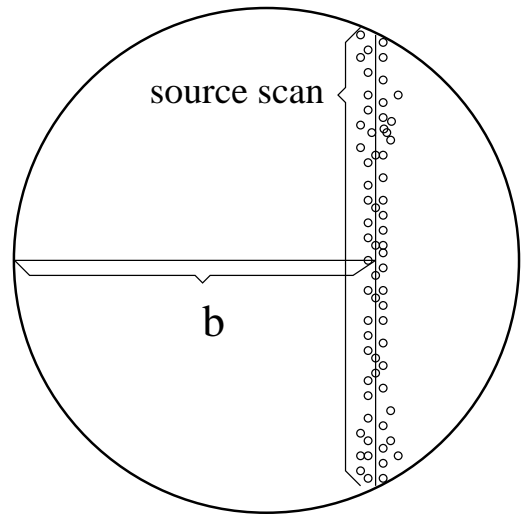}
      \caption{Schematic view of a single scan of a source in the
               PSPC detector and its
               impact parameter $b$. The big circle indicates the
               detector field of view, while the little circles
               represent the individual photons detected from the given source.
               The vertical line is the calculated apparent path of
               the source through the detector. The actually recorded photons
               scatter around this regression line due to the detector's
               point response function.
              }
         \label{impact}
   \end{figure}

The RASS observations were carried out between 1990 July 30 and 1991
January 25; a few gaps in the survey were filled in July 1990, in February and
in August 1991.
During the RASS the satellite scanned the sky in great circles.
Scanning and orbital period were the same with each orbit lasting
96\,min.  The scans circles were perpendicular to the plane of the
ecliptic and contained the ecliptical poles. The ecliptical longitude of the
instantaneous scan longitude moved approximately with the apparent speed of the Sun
along the ecliptic; thus the whole sky was covered within
half a year. Since the circular field of view had a
radius of 57\,arcmin (and the angular velocity of the Sun is $\sim $ 1\,degree per day)
a source near the ecliptic plane was
scanned for about two
days while sources at the ecliptical poles were scanned during
the whole survey. The period $T$ during which that a given source is observed
is a function of the ecliptical latitude $\beta$ and is computed as
\begin{equation}
T \sim \frac{2}{\cos(\beta)}.
\end{equation}
The length of a single scan can last up to  30.4\,s if the
source passes exactly through the center of the field of view; note
that the scan speed  was 3.75\,arcmin/s.
The exact length of a scan depends on the impact parameter b, which is
defined as the nearest distance of a given source to the left (western) border
of the detector during a given scan; the scan geometry is illustrated in
Fig. \ref{impact}.

Clearly, the RASS data for each source consist of
a series of snapshots, each of which
has a different exposure time (depending on its impact parameter).   The
individual snapshots follow each other with the orbital period of
of $\approx$  96\,min.


\section{Data Analysis}
\subsection{Source detection}

A meaningful search for variability is possible only for
sufficiently strong sources.  For example, a source with
0.1 cts/sec will produce only three expected counts in the
longest possible scan with an signal-to-noise ratio (SNR)
of less than two.  In order to find sufficiently strong sources
we proceeded as follows:  A source detection was first carried
out on the merged survey data
with a maximum likelihood algorithm; an X-ray  source was
accepted for further analysis if it exceeds
the threshold of the maximum likelihood value of 15
(which corresponds roughly to 5 sigma). This resulted in about 30\,000 
point sources for which lightcurves were generated.
For the source detection and
generation of these lightcurves the EXSAS context within
MIDAS (distributed by the European Southern Observatory) was used; for
description of the EXSAS environment see \citet{EXSAS}.

\subsection{Extraction of lightcurves}

For the construction of X-ray lightcurves we extracted all photons
within a circular region centered on the nominal source position.
The correct choice of the size of the extraction region is a
nontrivial problem.  Clearly, the larger the extraction region the more
background is picked up; on the other hand, because of the sensitivity
of the point response function with off axis angle, too small an extraction
region leads to an unacceptable loss of source photons.  Ideally one should
work with a "breathing" detect cell that changes its size according to
the actual off axis angle, a process computationally a bit cumbersome.
Since meaningful results are expected to emerge only from stronger sources,
we decided instead to work with two different extraction radii depending
on the scan impact parameter $b$,  one for the central scans with middle
 values of
$b$ and one for the non-central scans with larger and smaller values of $b$.
The detect
cell sizes were determined in such a way that 90 \% of the source
photons must be collected; the resulting lightcurves were then corrected
for the missing source photons.

\begin{figure}
   \centering
   \includegraphics[width=7cm,height=6cm]{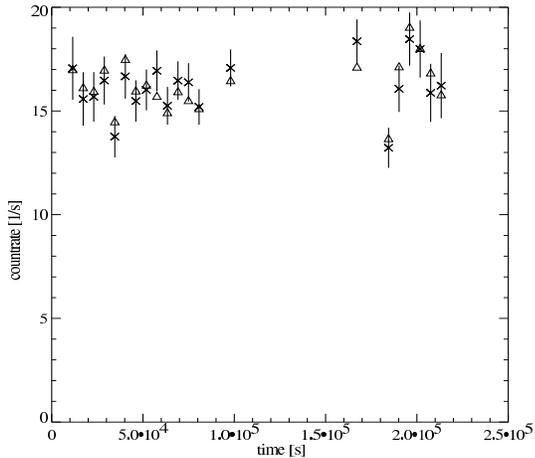}
      \caption{Comparison between the two extraction methods applied to the
               strong white dwarf source Sirius B. The crosses with error bars
               indicate the count rates computed using a fixed radius detect
               cell large enough to collect more than 99 percent of the
               source photons. The triangles indicate the values from our
               adopted two radii method with a correction to account for the
               full photon flux.
              }
         \label{chitestsir}
   \end{figure}

We tested this method using four strong sources, i.\,e. Capella,
Algol and the white dwarfs Sirius B and HZ 43; especially the latter
two sources are expected to be constant.  All of these sources are
so strong that one can choose a fixed radius which collects on average more
than 99 percent of the source photons while the background level
is negligible.  A comparison of the lightcurves generated with
the two methods resulted in differences at a level
of about 5 percent (see Fig \ref{chitestsir}).
Since the variability effects that can be meaningfully searched for in the RASS data
are much larger we conclude that our method is acceptable and does not introduce
any artificial variability.  Also note that scan to scan variations of up to
20 \% can occur for presumably constant sources; any variability below that level may
therefore be spurious.

\begin{figure}
   \centering
    \includegraphics{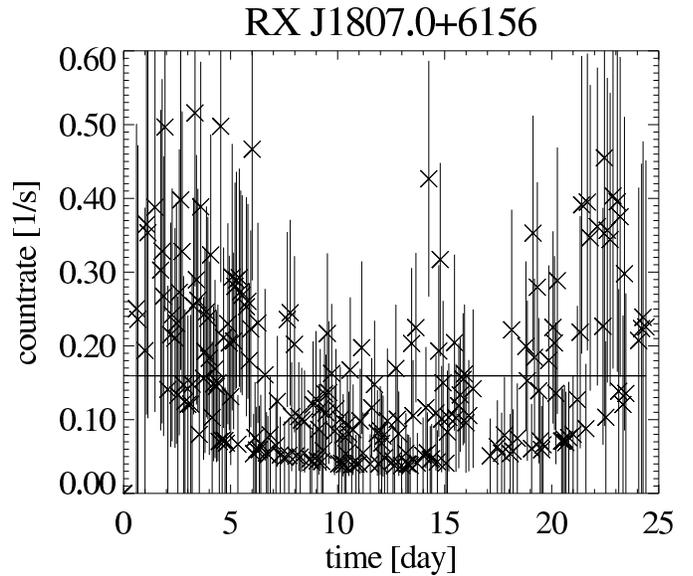} 
      \caption{Example of a RASS lightcurve showing a typical curvature
               with the count rate rising towards both start and end of the
               observations; note that
               scans with zero counts and zero exposures are not included.
              }
         \label{bent}
   \end{figure}

For the generation of the lightcurves we used the EXSAS procedure
\texttt{create/accepted}, which computes the exposure time per scan from
the known scan speed and the average photon impact parameter $b$ of the
scan. Since for many scans there are only a few photons defining the
scan path on the detector plane,
this method is not very exact. To improve the accuracy of the impact parameter the procedure
automatically performs a linear fit of $b$ against the average photon arrival
time $T$ of each scan. The relation between $b$ and $T$ is theoretically
an arc of a sinusoid, but a straight line approximation is good enough
for all sources with a distance of more than 2 degrees
from the ecliptic
poles. In the immediate vicinity of the ecliptic poles the approximation of a
straight line breaks down, because the sources are detected in every scan.
Hence the center position of the source in each scan is the projection
of the circular motion of the source onto the detector and therefore a
sinusoid.  A second reason led us to
excluded these sources from our analysis.
All these sources turned out to be very weak and for such sources
at a distance of less than
about 15 degrees of the ecliptical poles there is another problem
illustrated in Fig. \ref{bent}.

Sources with an average count rate of less than $\sim$0.5 photons per second
show a typical curvature that increases with decreasing count rate.  This
curvature is not caused by an incorrect vignetting correction, rather
it is due to the survey geometry and the small number counting statistics:
The exposure time in the scans near the edge of the detector and therefore at
beginning and end of the survey observations is very short. If a single photon or
maybe two are detected in one of these scans,  the count rate is consequently high.
If, on the other hand, no photon is detected, no impact parameter and
no data points can be derived with our method.  Near the ecliptic poles the impact parameter
b changes only slowly from scan to scan thus causing
the described curvature of the lightcurves. In principle, the same
effect is also present for sources at lower ecliptic latitudes but it does not
become apparent because of the rapid change of impact parameter b with increasing
scan number.  In order to confirm this interpretation
we simulated lightcurves where the photons  for each scan were
computed assuming a fixed count rate under Poisson noise. For
these artificial sources lightcurves were generated showing the same
curvature as observed in the ROSAT lightcurves, too.  The curvature could be removed
by averaging over several scans at the expense of temporal resolution.  However,
since we have no easy way to distinguish between scans with zero counts and scans with zero
exposure we decided not to pursue this matter any further.
Unfortunately the lightcurves with curvature
turned out to be a major problem for our variability search algorithms, since both spurious
flares and periods were recognised by our search algorithms.

\subsection{Search for flares}

For the search for flares we developed an algorithm  combining different
tests. Altogether  ten tests were applied, and if five give a positive answer
then the lightcurve is assumed to be variable. Out of the ten tests seven are
empirical and three are statistical ones. The statistical ones are the
Kolmogorov-Smirnov test and the $\chi^{2}$-test with a double weight. For
a description of these two tests see e.\,g. \citet{Press}. 
For the Kolmogorov-Smirnov test a false-alarm probability
threshhold of 40 percent was used. This high threshhold can be used because
the KS-test is only one indicator for a flare out of ten. Using a lower
threshhold makes the KS-test insensitive to minor flares. The $\chi^{2}$-test
was adapted using 
\[\chi^{2}=\frac{1}{n}\cdot\sum_{i=1}^{n}\frac{(\mu-r_{i})^{2}}{e_{i}^{2}}\]
with $\mu$ being the mean, $r_{i}$  being the count rate in the i-th scan
and $e_{i}$ being the error of the count rate. If $\chi^{2} > 1.8$ then
the scan with the largest deviation is searched and excluded. A new $\chi^{2}$
is computed and the process is iterated until $\chi^{2} < 1.5$. If one
of the excluded scans has a count rate higher than the mean, then the test
is positive. If the excluded scans have all count rates lower than the mean
the lightcurve is not thought to be variable since dips can be instrumental.
The two threshholds for $\chi^{2}$ were empirically determined with a
 testfield.
The empirical
tests try to identify signatures in the lightcurves which have the typical
shape of a flare. Specifically, we searched for a time bin with a count rate
three times higher than the median, for four time bins
following each other, each higher than the median and each one lower than the
prior one.

We tested our search algorithm with simulated lightcurves. First we created
constant lightcurves with only Poisson noise imposed and tested the number
of lightcurves associated with spurious variability. We realized - not
surprisingly -
that this kind of error depends on the mean count rate. With a high count rate
(10 counts per second) about 4 percent of the simulated lightcurves are
classified as variable, while with a low count rate ($<0.3$ counts per second)
about 0.3 percent is classified as variable despite their constancy.
 The bent lightcurves we found
(see \ref{eye}) were not taken into account in this test.

\subsection{Search for periods}

The search for periods was performed with the  Lomb-Scargle Algorithm
(see \citet{Press}), that can deal with discrete, unevenly spaced data.
The algorithm computes a periodogram, in which the amplitude translates
directly into a false-alarm probability of the apparent period. The
threshold for the false-alarm probability for variable sources was set
to 5 percent. Moreover the found period must not be longer than 95 percent
of the lightcurve to be accepted.

\subsection{Search for trends}

In lightcurves with ecliptic latitude $\beta\gppr|60|^{\circ}$ corresponding to
a minimum of four days observation was also searched for trends. The algorithm
compared the median of the first quarter of the lightcurve with the median
of the last quarter of the lightcurve. If the difference exceeds 30 percent
of the median of the whole lightcurve and is at least 0.2 then a trend is
assumed.

The sources with ecliptical latitude $\beta\lppr|60|^{\circ}$ were not
searched for trends since the lightcurves consist of only few scans so that the
the median of the lightcurve becomes less robust.

\subsection{Revision by eye}\label{eye}

The analysis of the lightcurves was performed in two steps. First, all lightcurves
with ecliptical latitude $\beta\gppr|60|^{\circ}$ were analysed, then the much shorter
ones at lower latitudes.  We found it necessary to visually inspect
all high latitude lightcurves because of the curvature feature described above.
The ill defined background level could produce spurious flares and the curvature
in the high latitude lightcurves
would produce spurious periods. Therefore
spurious variability arising from such curved lightcurves was
removed manually.  The data quality of
lightcurves at low ecliptical latitude turned out to be much better and we did not find
it necessary inspect all the lightcurves  by eye.
The number of variable sources found at both low and high ecliptic latitudes
is given in Table \ref{num}.  Table \ref{num} shows very clearly that at
low ecliptic latitudes flares are detected almost exclusively; at high latitude
this pattern changes completely, however, we also found it necessary to reject
a significant number of sources that probably show only spurious variability.

   \begin{table}
      \caption[]{Number of variable sources found in the RASS. In brackets
                 the number of sources researched by eye. In the lower part
                 are the numbers of sources which show two types of
                 variability. }
         \label{num}
         \begin{center}
         \begin{tabular}[htb]{cccc}
            \hline
            \noalign{\smallskip}
            \multicolumn{3}{c}{number of sources analysed} & 29970   \\
            \hline
             & $ \beta \gppr |60|^{\circ}$ & $ \beta \lppr  |60|^{\circ}$ &total\\
            flares & 330 (664) & 664 & 994\\
            periods & 85 (887) & 60 & 145\\
            trends & 90 (412) & - & 90\\
            \noalign{\smallskip}
            \hline
            flares and trends & 2 \\
            periods and trends & 2 \\
            flares and periods & 18 \\
            \hline
         \end{tabular}
         \end{center}
   \end{table}
%


\section{Optical identification of the variable sources}

In order to perform a statistical analysis of the variable sources it is
necessary to identify them with optical counterparts. For this purpose
we used three catalogs, i.\,e. the SIMBAD database, the
USNO-A2.0 catalog and the NED. SIMBAD (Set of Identifications,
Measurements, and Bibliography for Astronomical Data) contains names, positions
and known properties of 3\,063\,000 objects outside the solar system that are
mainly galactic.
The USNO-A2.0 catalog of the U.S.
Naval Observatory contains 526\,280\,881 stars with positions and
brightness in the R and B band. The NED (NASA/IPAC Extragalactic Database)
contains names, positions and basic data for about 4\,700\,000 extragalactic
objects.

\subsection{Identification with the SIMBAD database}

Since we expect mostly stars to show variability (flares and periods) we
first tried to identify the detected variable sources with
counterparts cataloged in the SIMBAD database.  In order to match
RASS sources with SIMBAD entries
we used a search radius of 1.5 arcmin which significantly exceeds
the error circle of 90\,\% of 30$^{\prime\prime}$ determined by
\citet{Krautter}.  Despite this rather generous search radius
many X-ray sources turned out to have
no cataloged SIMBAD counterpart.

\subsection{Identification with the USNO-A2.0 catalog}

For the sources that remained unidentified with SIMBAD entries
or that were only listed as X-ray sources in SIMBAD we tried to find
counterparts using the USNO-A2.0 catalog.  In order to limit the number of counterparts
to a reasonable level
we constrained the search radius to 40$^{\prime\prime}$, a procedure that
left us with at least one candidate object for each source.  In order
to assess the plausibility of this matching procedure by
positional coincidence
we estimated the ratio of the X-ray flux
to the R band flux and defined a threshold above which the
candidate stars were not accepted. The X-ray flux was estimated from
mean count rate using the conversion factor $\alpha = 6 \cdot
10^{-12} \frac{\mathrm{erg}}{\mathrm{counts\ }\,\mathrm{cm}^{2}}$.
 The flux $f_{\lambda}(R)$
in the R band outside
the atmosphere was computed from $\log\,f_{\lambda}(R)=-0.4m_{\mathrm{R}}-8.58 +3$
where $m_{\mathrm{R}}$ is the brightness in the R band; the width of the R band was
assumed to be 1000\,\AA. The value 8.58 applies only for B stars and changes
slightly with the spectral class; the exact numbers can be found in
\citep{Allen}.  Note that for O stars the equation is not valid; however, since
variability among O-type stars is quite unusual this should hardly matter.

In order to define sensitive thresholds we used typical maximal values
of this ratio known in the V band from \citet{Krautter} and of our own
sample that was identified with SIMBAD. In our own sample we find a
tendency to higher values for both the lower and the upper limits of the
ratio which we ascribe to flares in this sample that increment the X-ray flux.
The determined values of the ratio in the V band as well as the used threshold
in the R band are listed in \mbox{Table \ref{XVratio}.}

Using these threshold values out of 436 X-ray sources not
cataloged as SIMBAD entry 183 X-ray sources
ended up with exactly one acceptable object within the search radius
while 149 objects had no acceptable entry within the search radius.
In all other cases more than one acceptable star was found in the search
radius so no unique (positional) identification can be performed.

 \begin{table}
\begin{center}
\begin{tabular}[htb]{ccc|cc|c}

\hline
Object class & \multicolumn{2}{c}{$\log(\frac{f_{X}}{f_{V}})$}&\multicolumn{2}{c}{$\log(\frac{f_{X}}{f_{V}})$}& used\\
 & min & max & min & max& threshold\\
\hline

B stars & -3.09 & -0.56 & - & - & -0.56\\
A stars & -3.58 & -3.41 & -4.59 & -3.78 & -3.41\\
F stars & -4.19 & -1.92 & -4.82 & -1.97 & -1.92\\
G stars & -4.37 & -1.84 & -5.52 & -0.80 & -0.80\\
K stars & -3.52 & -0.10 & -4.57 & +0.20 & +0.20\\
M stars & -1.99 & +1.30 & -4.41 & -0.14 & -0.14\\

\hline

\end{tabular}
\caption{At the left are shown the limits of the X-ray to V band flux ratio
found in the identified sample. In the middle the same for the sample
\citet{Krautter} used. At the right the used threshold for the R band.}
\label{XVratio}
\end{center}
\end{table}

\subsection{Uncataloged objects}

\begin{figure}
   \centering
   \includegraphics{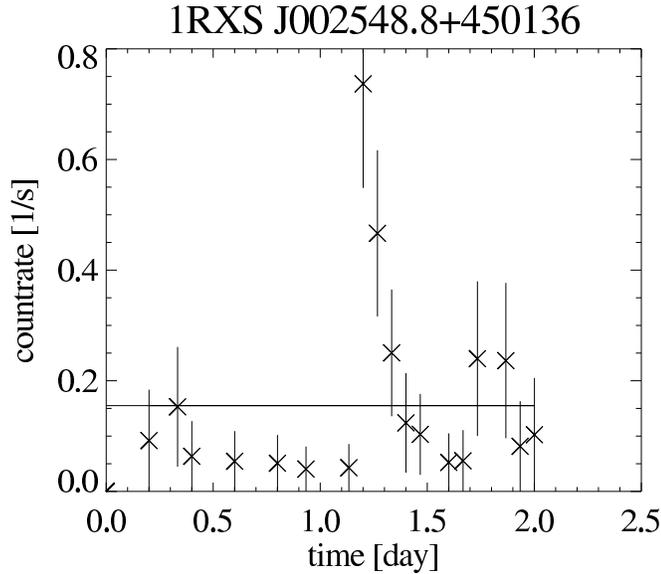}
      \caption{Example of a typical lightcurve of an optical not cataloged
               source. Note that strong flare with its sudden rising and
               long lasting decay strongly suggesting a stellar nature of
               this event.
              }
         \label{short1}
   \end{figure}

Since the number of objects with no stellar
counterpart in the USNO catalog is quite high these objects must either be
very active in the X-ray band or very faint in the R band or they are no stars
at all.  In order test this hypothesis
we used the NED data base. For the
149 objects without stellar identification we could identify 15 with
galaxies and 4 with quasars listed in the NED data base but not in SIMBAD;
the remaining 130 sources are still unidentified. We suggest that these
are likely to be stars for the most part due
to two reasons. First there are stellar counterparts for all sources found
in the USNO-A2.0 catalog without applying the threshold ratio of the X-ray
to visual flux. Taking into account that most of the unidentified
objects show flares the threshold is presumably too low and therefore
the stellar content of the sample is underestimated. The second reason is
implied by the lightcurves themselves: Many of the unidentified sources do
in fact show typical stellar flare lightcurves as shown in Fig \ref{short1}; it
would be a conspiracy of nature if the counterparts were not stars.

In Table \ref{ident} we give a summary of the
identifications obtained from the SIMBAD, USNO-A2.0  and NED data bases.
The variable objects are distinguished according to their stellar types as well
as to the type of variability found; note that several types of variability
can be found for the same lightcurve.  The quoted percentages refer to the
percentage of the respective source class among all the variable sources.
As is obvious from Table \ref{ident}, late-type stars are responsible for the
largest portion of the variable source population; about 10 \% of the
variable sources are of extragalactic origin.

A list of all ROSAT all-sky survey variable sources is given in Table
\ref{variability}, avaiable at the CDS, where we list
the name, position, type and the luminosity (if known) of the objects.
Furthermore there are variability flags given showing which sort of variability
was found. The flags indicate the following: fa - flare automatically found,
f - flare, pa - period automatically found, p - period, ta - trend 
automatically found, t - trend. 

 \begin{table}
\begin{center}
\begin{tabular}[htb]{ccccc}

\hline
Object class & total & flares & periods & trends\\
\hline

Sy1 & 60 (5.0) & 44  & 6 & 12\\
Sy2 &  3 (0.2) & 0  & 1 &  2 \\
AGN & 36 (3.0) & 24 & 5 &  7 \\
Gal & 19 (1.6) & 10 & 3 &  6 \\
 & \\
LXB & 8 (0.7)  & 7  & 0 &  1 \\
SNR & 2 (0.2)  & 1  & 0 &  1 \\
 & \\
WD & 13 (1.1)  & 9  & 5  & 1 \\
O & 2 (0.2)    & 2  & 0  & 0 \\
B & 9 (0.7)    & 8  & 2  & 1 \\
A & 12 (1.0)   & 10 & 2  & 0 \\
F & 45 (3.7)   & 40 & 3  & 2 \\
G & 95 (7.9)   & 77 & 13 & 7 \\
K & 239 (19.8) & 205& 27 & 10\\
M & 137 (11.3) & 134& 2  & 1  \\
star & 215 (17.8)& 191 & 22 & 9 \\
 &  \\
not uniquely & \\
identified & 173 (14.3) & 144 & 22 & 8\\
not id with& \\
 cataloged obj & 139 (11.5) & 88 & 32 & 22\\
total & 1207 & 994 & 145 & 90\\

\hline

\end{tabular}
\caption{The identification of the variable sources. Given are the numbers
of objects of each object type for the different kinds of variability.
 In brackets is shown the percentage. AGN is
active galactic nuclei, LXB is low mass X-ray binary, SNR is supernova remnant,
WD is white dwarf. The sources in the category 'not identified with
cataloged objects' are mainly identified with X-ray
sources, with some minor identification of infrared and radio sources.
The objects in the category 'stars' are stars from the SIMBAD
database with unknown spectral type and stars from the USNO catalog where
more than one star with a sensible X-ray to optical flux ratio was found.
The sources not uniquely identified with SIMBAD
are found in the category 'not uniquely identified'.}
\label{ident}
\end{center}
\end{table}

\subsection{Unexpected identifications}

X-ray variability among late-type stars does not come unexpectedly.
However, there are some variable objects found in classes that are not expected
to show any variability at all. The former are the classes SNR and
WD, the latter are the classes of stars with spectral type O, B and possibly A.
A closer inspection of the individual sources shows the
two O-stars to be high mass X-ray binaries (HMXB) as well as 4 of the
B-stars (and one of the WDs); the specific sources are \mbox{V779 Cen},
\mbox{4U 2206+54}\,(O-stars), V662 Cas, LMC X-3, GP Vel, 2A 0116-737 (B-stars)
and Her X-1 (WD).  Further,
two of the B-stars are double or multiple stars, namely CCDM J20016+7027AB and
HD 5394, while 5 of the A-stars are
eclipsing binaries of Algol type (HD 18022, HD 139319, HD 153345,
\mbox{V752 Sqr}, HZ Dra). Some of the A-stars are
identified using only the USNO-A2.0 catalog and the threshold for the X-ray
to optical flux ratio; the spectral type is therefore quite uncertain.
The other early-type stellar sources are X-ray sources
listed in the SIMBAD database (HD 21364 and HD 33904 are B-type stars,
HD 43940 and HD 116160 are A-type stars, the nova-like star
is the CV BL Hyi)

All of the variable SNRs (SNR 021.0+63.0, SNR 0519-69.0 and SNR 0525-66.0)
and some of the WDs have high
count rates and therefore a relatively wide dispersion in the data which can trick
especially the flare search algorithm (cf., Fig. \ref{chitestsir}).
The four WDs for which
periods have been found will be discussed below. One WD has an M-type companion
(IN CMa), two others (WD 0809-728 and WD 1648+407)
show small and rather long term variability that
cannot be explained by high count rates or known physical properties.
Out of these two WD 1648+407 is suspected to be a
binary \citep{Green}, but this needs confirmation. We checked for
stellar counterparts of these objects in the USNO-A2.0 catalog. A search
with a 1.5 arcmin radius revealed in both cases nearby red stars, so these
two white dwarfs might be misidentifications.
Remarks on these objects can be found in Tab \ref{special}.
In addition to these object classes there are three objects for which in
SIMBAD only galaxy clusters or groups
are found as possible counterparts. On the
other hand, the X-ray lightcurves of at least two of these three
objects show flares and look in fact quite stellar. It was confirmed with the
USNO-A2.0 catalog that there are possible stellar counterparts for these
objects. We therefore placed these three objects in the category 'not uniquely identified'.

\begin{table}
\begin{center}
\begin{tabular}[htb]{ccc}
\hline
Object class & Number & Remarks\\
\hline
O-stars & 2 & 2 HMXBs\\
B-stars & 9 & 4 HMXBs\\
        &   & 2 variable X-ray sources\\
        &   & 2 double star\\
        &   & 1 SNR (high count rate)\\
A-stars & 12 & 5 eclipsing binaries\\
        &    & 1 Novastar\\
        &    & 4 with USNO identified\\
        &    & 2 known X-ray sources\\
SNR     & 2  & 2 high count rate\\
WD      & 13 & 5 high count rate\\
        &    & 1 HMXB\\
        &    & 4 periods (discussed later)\\
        &    & 1 binary\\
        &    & 2 possible misidentifications (see text) \\
\hline
\end{tabular}
\caption{Remarks on the objects for some peculiar classes.}
\label{special}
\end{center}
\end{table}







\section{Some examples of variability}

In addition to the statistical analysis of the variable sources we
present some highlights of variability detected in the RASS data; because
of our specific interest in stars and the broad variety of
variable X-ray sources this selection is subjective by necessity.
We focus on objects where the
detection of variability is mostly due to the long observation times of
the RASS, i.e., variability that would have probably remained undiscovered in
(shorter) pointed data.

\textit{Note: The solid line in all the lightcurve plots indicate the mean
count rate.}


\subsection{New X-ray flare stars}

\begin{figure}
   \centering
    \includegraphics{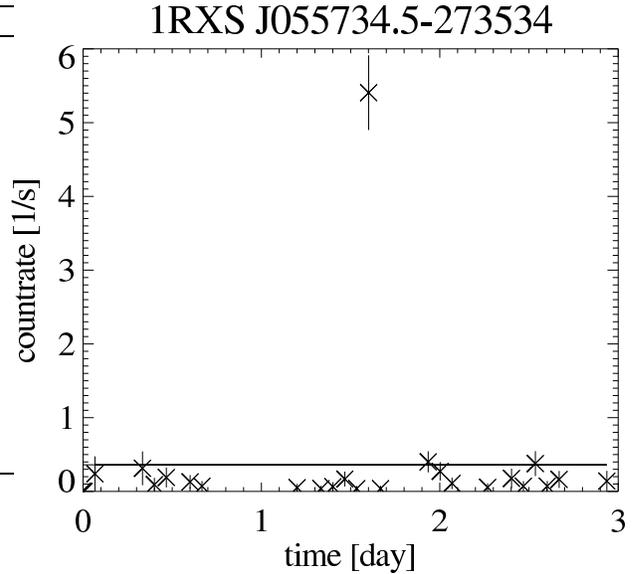}
      \caption{The RASS lightcurve for 1RXS J055734.5-273534
with a short duration flare detected only in one single scan.  Note the
enormous peak X-ray flux of this event.}
         \label{shortX}
   \end{figure}

Many X-ray flares were detected on objects that could either not be identified
with optical counterparts or could only be
identified with stars listed in the USNO-A2.0 catalog. Both
groups contain objects that show long duration flares with both the
flare onset and the whole decay observed as well
as objects that show sporadic short flares with very high apparent fluxes.
Some of these short duration events were already found by \citet{Greiner}
in their search for $\gamma$-ray bursts, but due to their decision of studying
only objects with a count rate consistent with zero before and after the
flare they in fact missed many actual flares (albeit no $\gamma$-ray bursters).
These short flares are often significantly detected  only in one time bin with
possibly more than one hundred photons in this bin. An example for such
an event on an essentially anonymous object without identification
in the USNO-A2.0 catalog is shown in Fig. \ref{shortX}.
The quiescent count rate is about 0.1-0.2 counts/sec rising to more than
5 counts/sec in the scan covering the flare.  From the available data it
is not possible to tell whether the data point is in the flare rise, peak or
decay.  The two scans following the flare are somewhat elevated, so that
this event could also be interpreted as along duration flare, but these
count rate enhancements are not significant.

\begin{figure}
   \centering
    \includegraphics{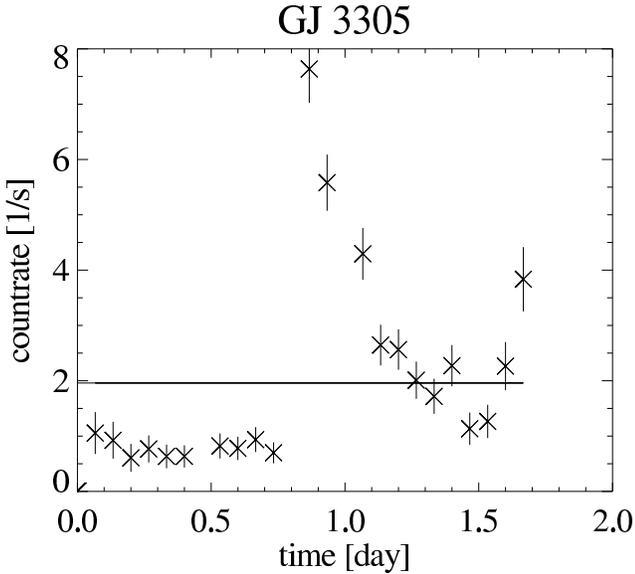}
      \caption{The lightcurve of GJ 3305. The long duration flare is fully
               observed including both onset and the whole decay.  Since the
               distance towards  GJ 3305 is known, the energetics of the
               flare can be calculated (cf. text for details).
              }
         \label{GJ3305}
   \end{figure}

There are more such outstanding flare events detected only in one scan,
but even more interesting are the newly detected long duration flares.
Some of these belong to previously known flare stars.
In Fig. \ref{GJ3305} we present the RASS lightcurve of the nearby
 (optical) flare star GJ 3305.  This star was observed twice during
the RASS, the first
time in the beginning of the all-sky survey showing no significant
variability, and second time at the end of the all-sky survey showing
this spectacular flare. Note that the flare decay lasts about one day
and produces yet another example of a long-duration flare on M-type stars.

We fitted the lightcurve of this flare with an exponential function
\begin{equation} c(t) = A_{0}e^{-\frac{t}{\tau}} + B \end{equation}
where $c(t)$ is the count rate as function of time $t$, $B$ the quiescent
background count rate, $\tau$ the flare decay time and $A_{0}$ the
count rate at flare peak.
The parameters  $\tau$ and $ A_{0}$ determine the radiated flare energy, which
is given by
\begin{equation} E = A_{0}\tau\alpha\cdot 4\pi R^{2} \end{equation}
where $R$ is the distance of the star and $\alpha$ the above mentioned
conversion factor. The two parameter $A_{0}$ and $\tau$ were fitted while
the background $B$ was fixed to the median.  From the best fit parameters
$A_{0}=7.4\pm1.1$ and $\tau=27633\pm8205$ and using the
distance of 15.2\,pc \citep{Jahreiss} we find a total released energy of
$3.4 \cdot 10^{34}$\,erg.  This number exceeds the typically quoted soft
X-ray energy releases of the largest solar flares by more than
two orders of magnitude, but compares well to other flares observed
on M-type stars, for example the long-duration flare on EV Lac described
by \citet{Schmitt1}.

\begin{figure}
   \centering
    \includegraphics{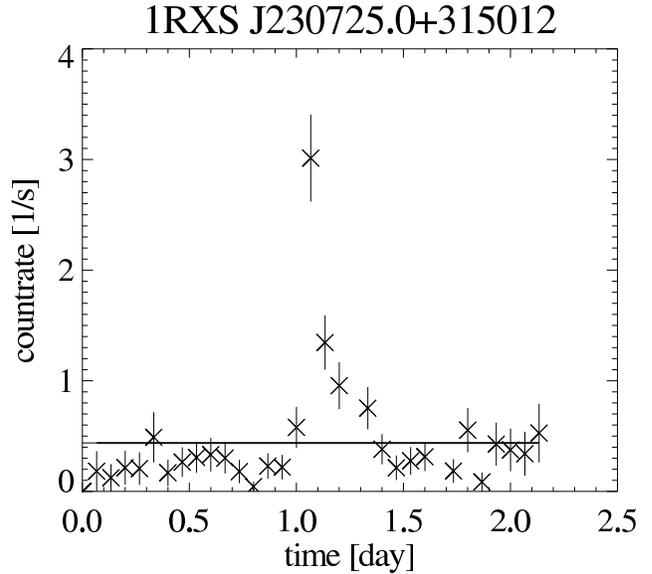}
      \caption{The RASS lightcurve of an anonymous star. It is tentatively
               identified with a K star listed in the USNO-A2.0 catalog.
              }
         \label{long1}
   \end{figure}

\begin{figure}
   \centering
     \includegraphics{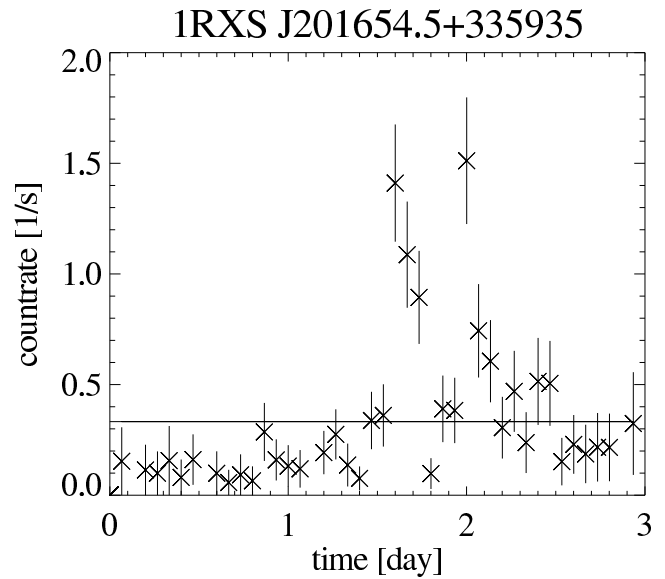}
     \caption{The RASS lightcurve of an anonymous star. It is not identified
              uniquely in the USNO-A2.0 catalog.  Note the multiple flare
              event which is quite typical for stellar sources.
              }
         \label{long2}
   \end{figure}

In addition to flares from previously known flare stars
we report the detection of long duration flares on previously unknown
flare stars with counterparts only in the USNO-A2.0 catalog. The lightcurves
of two such objects are shown
in Fig. \ref{long1} and Fig \ref{long2}.

\subsection{Periodic variability}

Periodic variability was found for a substantial number of
X-ray sources belonging to different source classes.  In addition
to periodic variability in short-period binary systems like RS CVns or
CVs, we also find periodic variability in sources where such variability
is uncommmon and hard to understand.  We specifically report
periodic variability for the galaxy RBS 1490 and for
four white dwarfs detected in our survey.
While periodic variability among active galaxies is rare and unusual,
quite number of the X-ray emitting WDs appear to be periodic.  The
four WDs in question are
all previously known white dwarfs and
all are very soft X-ray sources as expected for white dwarfs,
therefore a misidentification is unlikely.  The results of our
period searches for white dwarfs are
given in Table \ref{WDs}; the error of the period is computed from the
estimated FWHM of the peak in the periodogram.  Below we discuss the
sources individually.

 \begin{table}
\begin{center}
\begin{tabular}[htb]{lcc}
\hline
Name & Period [h]& fap\\
\hline
WD 1057+719 & $25.3 \pm 6.3$ & 2.1\\
GD 394 & $26.8 \pm 3.0 $& 1.2 \\
FBS 1500+752 & $11.6 \pm 0.3 $& 4.3\\
1ES 1631+78.1 & $69.4 \pm 8.3 $& 4.4 \\

\hline

\end{tabular}
\caption{Designations of the four periodically variable WDs, the RASS
determined periods and the false-alarm probability (fap) in percent.}
\label{WDs}
\end{center}
\end{table}

\subsubsection{The active binary CF Tuc}
\begin{figure}
   \centering
    \includegraphics{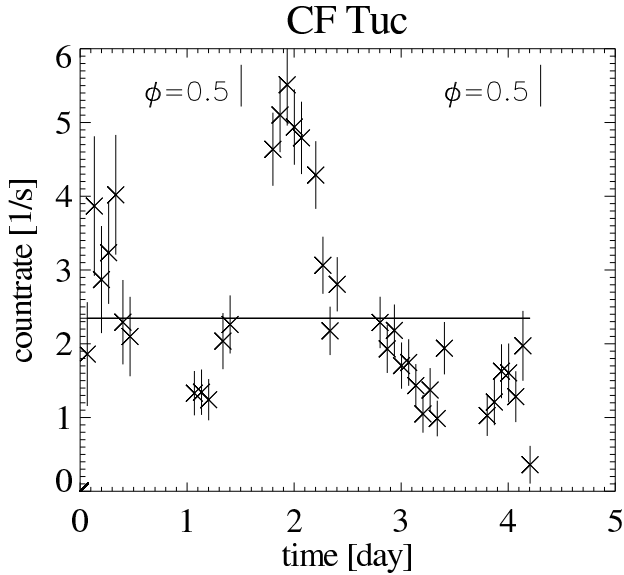}
      \caption{RASS lightcurve of the active binary CF Tuc.
              }
         \label{CF}
   \end{figure}

CF Tuc (HD 5303) is a partially eclipsing RSCVn system whose components are
of spectral type G0IV and K4IV \citep{Coates}. For this system an orbital
period of 2.798 day is known from UBV photometry \citep{Budding}, the
rotation of the two system components is
synchronised with the orbital period.
The RASS X-ray lightcurve of CF Tuc is shown in Fig. \ref{CF}; its high
degree of variability is self-evident.  Our X-ray period of $2.7\pm 0.8$
days is in full agreement with the optically determined period.
In Fig. \ref{CF} we also marked the times of optical secondary minimum
as computed from the ephemeres HJD 2\,444\,555.009 + 2.7977672d $\times$ E
\citep{Kuerster}.  Clearly, the X-ray minima do not coincide with any
optical minimum.  Alternatively, the RASS lightcurve of CF Tuc can also
be interpreted as a long duration flare with a duration length close to the
optical period; while this appears to be a strange coincidence, CF Tuc is known
to be capable of producing even much longer flares (cf., \citep{Kuerster})

\subsubsection{The cataclysmic variable EU UMa}

\begin{figure}
   \centering
    \includegraphics{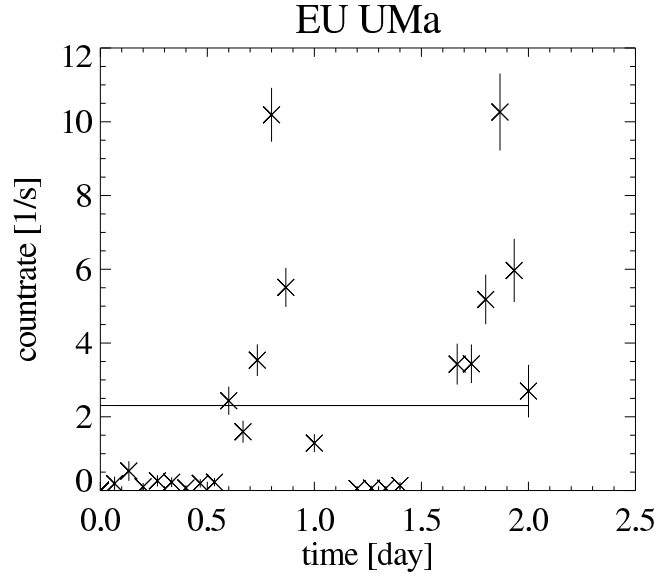}
      \caption{RASS lightcurve of the cataclysmic variable EU UMA.
              }
         \label{EU}
   \end{figure}

The RASS lightcurve of the cataclysmic variable of AM Her type EU UMa is
in Fig. \ref{EU}. One clearly sees a period of roughly one day (24.9 hours),
which is much larger than periods typically found for CVs and specifically
the EU UMa system period of 90.144 minutes \citep{Ritter}.  Following
\citep{Mittaz}, who present the ROSAT-WFC survey data for this source,
we interpret the observed period $P_{obs}$ as a result of folding the satellite
period of $P_{obs} = 96.2$ min with the binary period $P_{bin}$ according to the
rule $P_{bin}^{-1}=P_{sat}^{-1}\pm P_{obs}^{-1}$ with $ P_{sat}$.  Taking this aliasing
into account, the ROSAT survey data result in a binary period of
90.36 minutes, consistent with the WFC and optical data of EU UMa.

\subsubsection{The galaxy RBS 1490}

\begin{figure}
   \centering
    \includegraphics{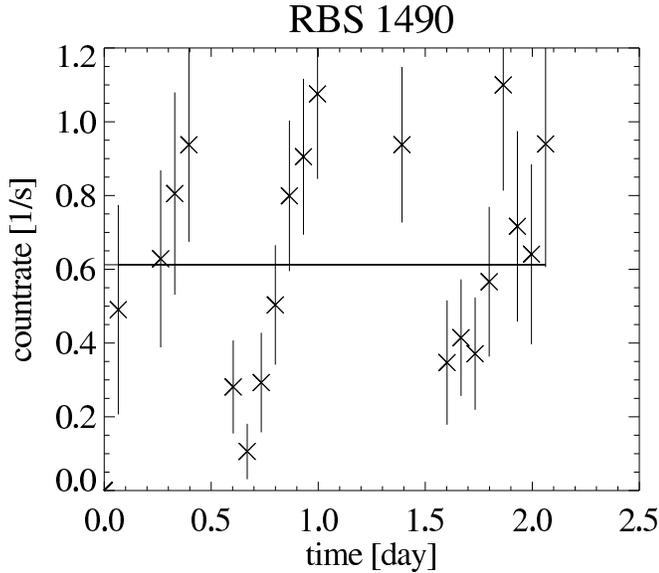}
      \caption{RASS lightcurve of the galaxy RBS 1490.
              }
         \label{RBS1490}
   \end{figure}

Another peculiar case of a RASS lightcurve is the case of the galaxy RBS 1490,
which shows a period of 12.2 hours; the lightcurve is shown in Fig. \ref{RBS1490}.
Since the variability amplitude is very high (more a factor of 5 variations from
peak to peak) the period is extremely significant, on the other hand, such a
temporal behaviour is not expected even for a Seyfert 1 galaxy. 
Therefore we checked for the presence of stars in the USNO-A2.0 catalog and found  four
stars in an area within a 1.5 arcmin search radius around the X-ray position. All
of those stars have too high a ratio of their X-ray to visual fluxes and we thus
reject those stars as possible counterparts.  We provide the
positions and provisional spectral types of those stars in Table \ref{RBSt} for further
investigation and conclude that at present no satisfactory explanation for the
RASS lightcurve of RBS 1490 can be given.


 \begin{table}
\begin{center}
\begin{tabular}[htb]{lcc}
\hline
Position & spectral type & $\log(\frac{f_{X}}{f_{V}})$\\

15 22 17.376  +16 48 28.39& A &0.39\\
15 22 15.519  +16 48 25.98& K &0.65\\
15 22 15.092  +16 49 16.63& F &1.72\\
15 22 13.462  +16 49 01.42& A &0.59\\

\hline

\end{tabular}
\caption{Positions (R.A. and $\delta$ (epoch 2000) and optical to X-ray flux ratios for the
four USNO stars in the vicinity of the source RBS1490.}
\label{RBSt}
\end{center}
\end{table}

\subsubsection{The white dwarf 1ES 1631+78.1}

The white dwarf 1ES 1631+78.1 is known to be a double system with
the binary being a dM4 \citep{Catalan}. The period found from the RASS
data is  \mbox{$69.4 \pm 8.3$\,hours} with a false-alarm probability of 4.4 percent.
This white dwarf (``meaty'')  was used as calibration
source for the Wide Field Camera (WFC) observing in the extreme ultraviolet
region during the ROSAT mission.  Besides a couple of
short pointed exposures there is also a 36 ksec calibration exposure
of this source. Unfortunately there is a data gap in this exposure and the
duration of the gap corresponds roughly to the found period. No periodic
variability is found in this lightcurve, but the length of the data gap
prevents us from ruling out the 69.4 hr period found in the survey data.

 \begin{figure}
   \centering
    \includegraphics{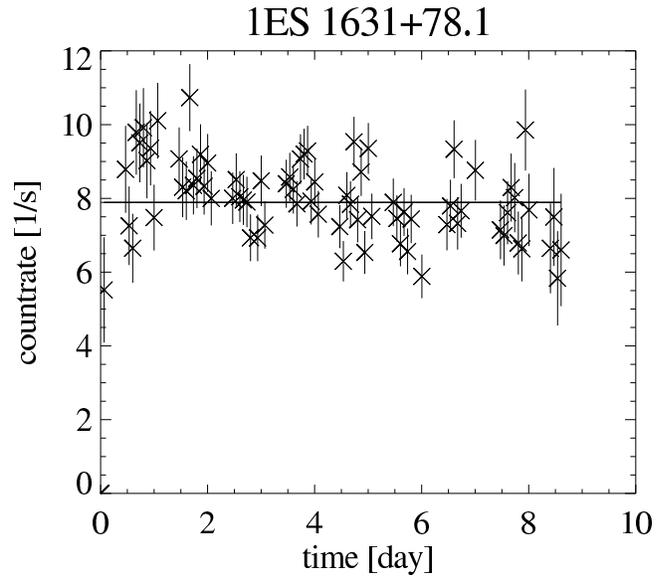}
      \caption{RASS lightcurve of the white dwarf 1ES 1631+78.1.
              }
         \label{1ES16}
   \end{figure}

\subsubsection{The white dwarf GD 394}

\begin{figure}
   \centering
    \includegraphics{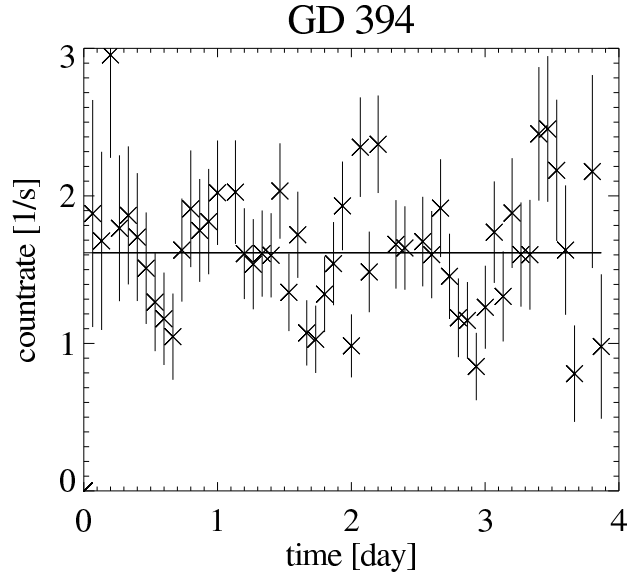}
      \caption{RASS lightcurve of the white dwarf GD 394.
              }
         \label{GD394}
   \end{figure}

The lightcurve of GD 394 is shown in Fig. \ref{GD394}; the period of 26.8 hours
is found with a false alarm probability of 1.2~\%.
The white dwarf GD 394 is also known to be variable in the EUV
although it seems to be single \citep{Dupuis}.
Analysing data from the EUVE satellite
Dupuis et al. found  a period of $1.150 \pm 0.003$
days, in perfect agreement with the period of $1.1 \pm 0.1$ days
we found in the RASS data.
Dupuis et al. interpret this period with a dark spot
on the surface of the WD. They discuss
two possible mechanisms that could produce such a spot:
First, a magnetic field and asymmetric accretion of material along this field
and second a comet may have been accreted. Since the data reported
by Dupuis et al. were taken in 1995 and the RASS data were taken in 1990,
we conclude that the spot has to be a
persistent feature with a life time of at least 5 years which argues against
interpretations invoking transient events.

\subsubsection{The white dwarf WD 1057+719}

\begin{figure}
   \centering
    \includegraphics{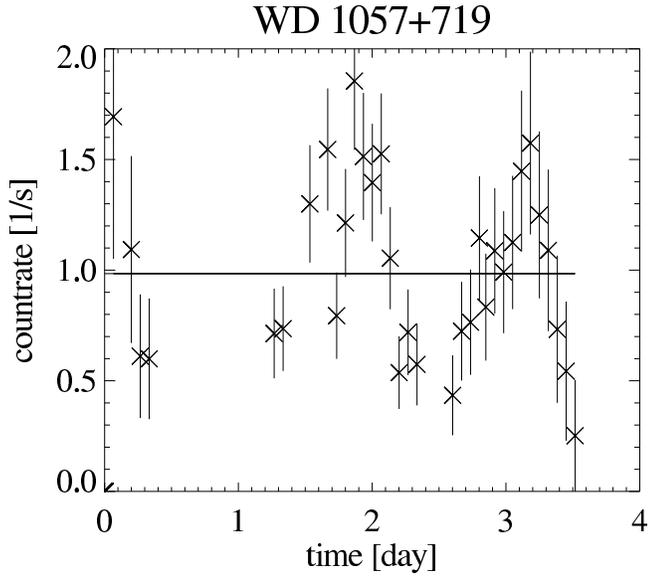}
      \caption{RASS lightcurve of the white dwarf WD1057+719.
              }
         \label{WD10}
   \end{figure}

Another periodically variable white dwarf is WD 1057+719; its lightcurve
and period ( $25.3 \pm 6.3 $hours) are strongly reminiscent of GD 394;
its lightcurve is shown in Fig. \ref{WD10}.  Like GD 394 WD 1057+719
seems to be single since \citet{Green} and we suggest it to be an analog
to GD 394.

\subsubsection{The white dwarf FBS 1500+752}

\begin{figure}
   \centering
    \includegraphics{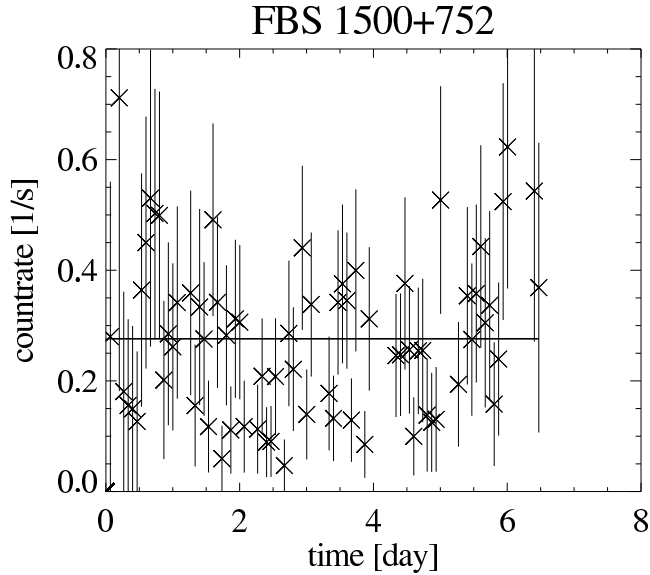}
      \caption{RASS lightcurve of the white dwarf FBS 1500+752.
              }
         \label{FBS15}
   \end{figure}

Very little is known about the object FBS 1500+752. It is cataloged as a white
dwarf in SIMBAD but no further information is available. We verified that the
source is supersoft. The lightcurve of this object can be seen in Fig.
\ref{FBS15}.  Our period search resulted in a period of 11.6 hours with
a false alarm probability of 4 \%.  Of the four white dwarfs with
periodic variability, FBS 1500+752 is by far the weakest and the lightcurve
is quite noisy, but definitely variable.  Further observations
are clearly needed to confirm the variability found in the RASS data.

\subsubsection{Periods not found: The case of HD48189}

\begin{figure}
   \centering
   \includegraphics{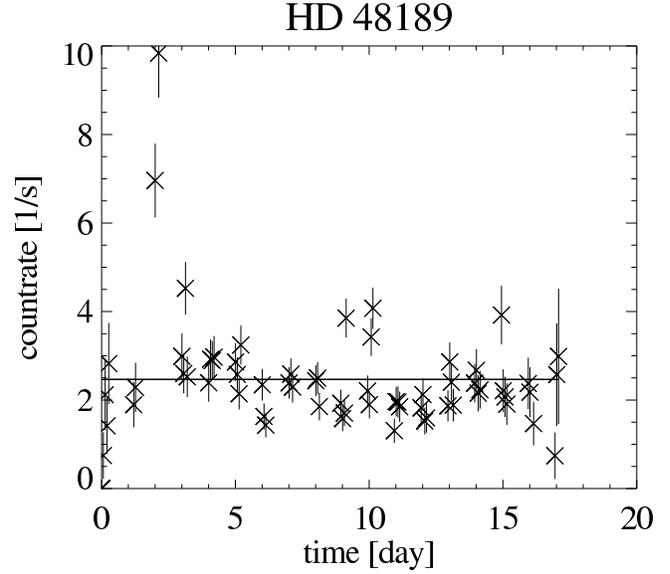}
      \caption{Lightcurve of HD 48189. Apparently there is a flare and
a period in the lightcurve.
              }
         \label{HD48189}
   \end{figure}

Our period search algorithm has problems finding periods in lightcurves
contaminated with flares or dips.  Consider the
the case of HD 48189, whose lightcurve is shown in Fig. \ref{HD48189}.
HD 48189 is a relatively nearby (its HIPPARCOS parallax corresponds to a distance
of 22 pc) visual binary with a lithium content and high X-ray luminosity 
(log L$_X$ = 10$^{29.8}$ erg/sec).  Both properties indicate youth.
The X-ray lightcurve of HD 48189 is characterised by two episodes of 
flaring which is not unexpected for alike stars.  Somewhat unusually,  
there also is
an underlying sinosoidal variation with a period of about 12.5 days in the
data. This period is not found by our algorithm with the full data set, however,
once the flares are manually removed, the period does show up significantly.
Of course, such a procedure is - strictly speaking - not legitimate; nevertheless,
it is very attractive to interpret the 12.5 day period as the rotation period of 
HD 48189.  However, \citet{Wichmann} determined a v sin(i) value of
17.0 km/sec from a high-resolution spectrum of HD 48189.  Assuming a solar
radius for HD 48189 we deduce a maximum period of 3 days in conflict with the
measured X-ray period of 12.5 days.  We thus conclude that the X-ray period
does not reflect the rotation period of HD 48189.  We note in passing that the
spectrum obtained by \citet{Wichmann} (available under web site http://www.hs.uni-hamburg.de/DE/For/Gal/Xgroup/projekte.html under topic young stars)
does not show any evidence for
binarity of HD 48189.

\subsection{Long term trends in active galactic nuclei}

\begin{figure}
   \centering
   \includegraphics{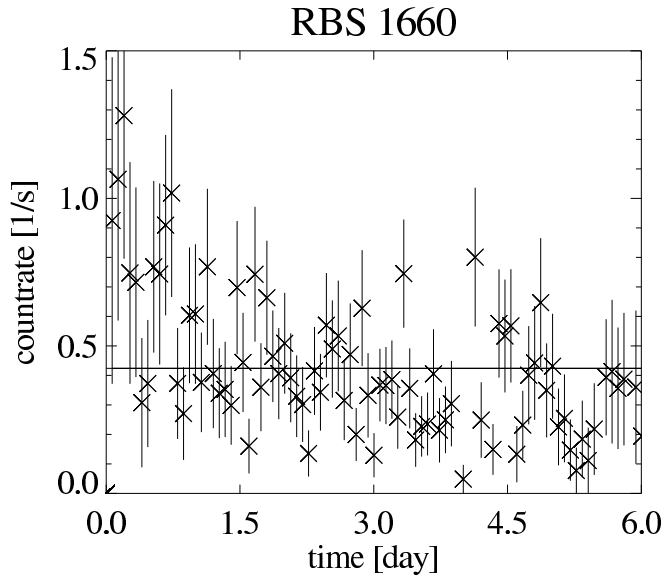}
      \caption{The RASS lightcurve of the Seyfert 1 galaxy showing besides
               minor short term variability a long term weak trend.
              }
         \label{Sy1}
   \end{figure}

\begin{figure}
   \centering
    \includegraphics{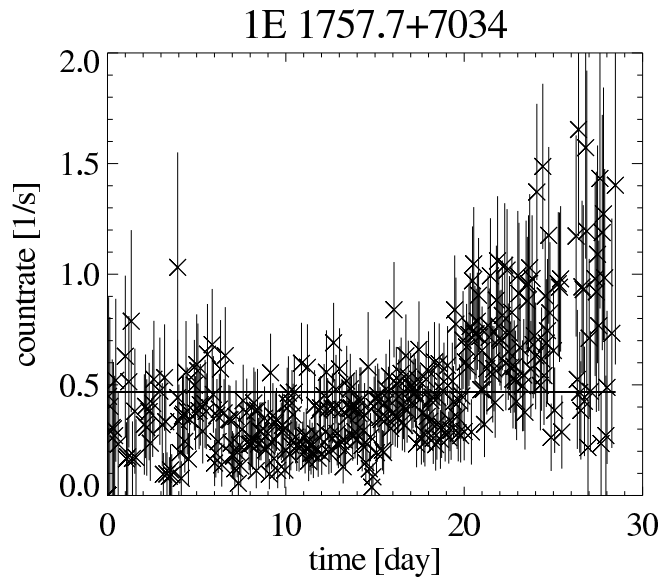}
      \caption{The RASS lightcurve of the BL Lac  showing a long term
                trend 1E1757.7+7034 with an increase in count rate of
                a factor of $\approx$ 3 over twenty days.
              }
         \label{bll}
   \end{figure}

Two galaxies with long trends were detected  that would not have been found in
considerably shorter observations. One is the Seyfert I galaxy RBS 1660
with a lightcurve covering about 7 days (see Fig. \ref{Sy1}); its
X-ray flux first decreases for about three days, after the minimum it exhibits
even shorter time-scale variability with a sharp increase by a factor of 2.
The other object is the well known BL Lac 1E1757.7+7034 with
lightcurve spanning 29 days shown in Fig. \ref{bll}.  For 1E1757.7+7034 the
X-ray flux stayed constant for about 2 weeks, after that it start a more
or less monotonic increase over the rest of the all-sky survey observations;
1E 1757.7+7034 is known to show intraday variability in the optical
\citep{Heidt} as well, but no X-ray variability has been reported so far.

%

\section{Discussion and conclusions}

Our study revealed widespread variability among the X-ray sources detected
in the RASS data.  Stellar sources account for the largest portion of the
variable source population. Among the stellar sources we found different
types of variability: High mass X-ray binaries with a high degree of
well studied variability but also flare stars, T Tauri stars
or cataclysmic variables. Unexpected variability behavior was, however,  also found:
Some active galactic nuclei showed a decrease or increase
in their count rate persisting throughout a couple of days.
Further, we found four white dwarfs showing periodical changes in their lightcurves
at high confidence levels.  One of these white dwarfs is known as a double system,
a second one has been very little studied so far, while the remaining two white dwarfs
seems to be single; the mechanism(s) producing a periodic modulation of the X-ray flux of
single white dwarfs can only be speculated upon.

Clearly the generous search radius used for our SIMBAD identifications
could introduce a bias toward galactic sources since SIMBAD contains more
galactic then extragalactic entries. In order to assess this possible bias
we checked a random sample of stars being identified with the help of SIMBAD
for extragalactic counterparts in the NED applying the same 
1.5 arcmin search radius
we used for SIMBAD identifications. We found that for at most 15 percent
of the SIMBAD identifications an alternative NED identification could be made
on positional grounds. However, since the probability for a variable
X-ray source being a star is very high 
we conclude that the number
of incorrect identifications of our variable X-ray sources must be small.

Interestingly and to us surprisingly, many of the variable X-ray
could not be identified with SIMBAD cataloged sources; these sources had therefore
not been identified as ``interesting'' (from their variability behavior) so far.
To overcome this problem we attempted to identify these X-ray
sources unknown to SIMBAD with stars from the USNO-A2.0 catalog using
sensible thresholds for the
X-ray to visual flux ratio. The main problem with this
approach is the optimal choice of the threshold.  We applied a very conservative
threshold with the consequence that
especially for flaring M-stars it is likely that the stellar
content of the unidentified sample is underestimated.

The mean magnitude of those stars identified with USNO catalog entries
is about 16 mag in the B-band and about 14 mag in the R-band.
Other than their apparent luminosity very little is known
about those faint stars, in particular there is no information on distances
either from parallaxes or from spectroscopy; this is unfortunate because it
prevents us from accurately assessing the energetics of the observed flares.
The peak fluxes of some of the detected flares suggest very high energies,
but distances are needed to confirm this suspicion.
For some of the brighter stars HIPPARCHOS parallaxes are known. For those stars we
computed the flare energy release from the fitted flare decay lightcurves and
found energy releases between $10^{33}$ and $10^{35}$ erg. Forthcoming astrometric
missions like DIVA or GAIA will provide distances for many of the stars identified
with USNO entries, eventually allowing to compute the released energies.

\begin{figure}
   \centering
    \includegraphics[width=7cm]{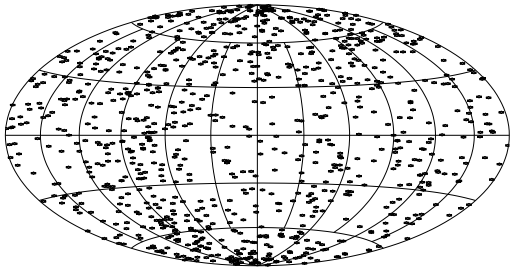}
      \caption{Aitoff projection of the flare sources on the sky drawn in
               ecliptical coordinates; note the concentration towards the poles
               of the ecliptic which can be ascribed to the RASS survey geometry.
              }
         \label{lb}
   \end{figure}

An Aitoff projection of the flare X-ray sources in ecliptical coordinates
supports the view that most of these sources are nearby stars (cf., Fig. \ref{lb}).
The surface distribution is relatively uniform with an increase of sources
towards the ecliptical poles. This is obviously a selection effect: Since the
observation times increase for sources near the ecliptical poles,
the chance for finding variability increases as well.

A natural question arising with these faint sources is their nature and age.
Since these stars are quite active as evidenced by their strong flares we
suspect them to be also rather young, yet no
obvious clustering of the flare sources (e.g. in open clusters, star forming regions etc.)
is observed (cf., Fig. \ref{lb}).  Apparently these stars are field stars of an young,
but otherwise unspecified age.  Possibly they represent the lower mass extension of
the young field star population studied by \citet{Wichmann}.

In addition to many flares on hitherto unrecognized flare stars we also find periodical
variability on some stars.  In a few cases these periods can be traced back to the
binary periods in known binary systems.   In other systems the periods are new
and might be due to rotational modulation caused by active regions on the 
surface. The periods found range between 10 and 770 hours
although we have searched for periods up to 1000 hours; the mean period is
about 130 hours which is certainly consistent with the interpretation as
rotational modulation.  We nevertheless have to state that it is rather difficult
to find rotationally modulated signals in the X-ray emission of stars except for known
eclipsing systems.  Since active stars, which the RASS data is most sensitive to,
first, show a lot of intrinsic variability and, second, have active regions which
may be predominantly concentrated near the stellar poles, 
rotationally modulated signals are expected to be weak and washed out by
stochastic variability.

%

\begin{acknowledgements}

We have made extensive use of the ROSAT Data Archive of the Max-Planck-Institut
f\"ur extraterrestrische Physik  (MPE) at Garching, Germany.
We particularly thank Rainer Gruber at MPE for technical support
with EXSAS.  This research has made extensive use of the 
SIMBAD database, operated at CDS,
  Strasbourg, France. This research has made use of the NASA/IPAC
Extragalactic Database (NED) which is operated by the Jet Propulsion
Laboratory, California Institute of Technology,
under contract with the National Aeronautics and Space Administration.
This research has made use of the USNO-A2.0 catalog using the VizieR
search page \citep{Vizier}.

\end{acknowledgements}

\bibliographystyle{aa}
\bibliography{H4169}


\onecolumn
\begin{landscape}
\scriptsize
\begin{longtable}{*{14}{c}}
\caption{Catalog off variable sources}\label{variability}
\\ \hline
   & \multicolumn{2}{c}{position in X-ray}&\multicolumn{6}{c}{flags for variability}
& & & & & \\
  Num & $\alpha_{\mathrm{2000}}$ & $\delta_{\mathrm{2000}}$ & f & fa &
  p & pa & t & ta & type & name & B & V & remarks \\
  & [h m s]& [° $^{\prime}$ $^{\prime\prime}$] & \multicolumn{8}{c}{ }& [mag]& [mag] & \\ \hline \\
\endfirsthead
 \multicolumn{14}{l}{Continuation of Table \ref{variability}}\\
 \hline
   & \multicolumn{2}{c}{postion in X-ray}&\multicolumn{6}{c}{flags for variability}& & & & & \\
  Num & $\alpha_{\mathrm{2000}}$ & $\delta_{\mathrm{2000}}$ & f & fa &
  p & pa & t & ta & type & name & B & V & remarks \\
  & [h m s]& [° $^{\prime}$ $^{\prime\prime}$] & \multicolumn{8}{c}{ }& [mag]& [mag] & \\ \hline
\endhead
\endfoot
\endlastfoot

0001&00 10 31.14&+10 58 34.5&1&1&0&0&0&0&Sy1&QSO B0007+107&0.0&0.0&            \\
0002&00 11 24.67&-11 28 38.5&1&1&0&0&0&0&star&V*WW Cet&9.3&9.3&dwarfnova\\
0003&00 12 34.60&-04 16 30.1&0&0&1&1&0&0&starU&1RXS J001234.8-041633&0.0&0.0&            \\
0004&00 13 09.50&+05 35 53.7&1&1&0&0&0&0&K4V&V*DV Psc&11.79&10.59&eclbinary\\
0005&00 13 55.33&-74 41 23.8&1&1&0&0&0&0&G6V&HD987&0.0&0.0&double*\\
0006&00 15 04.18&-72 50 28.9&1&1&0&0&0&0&M2V&EXO 001239-7307.3&0.0&0.0&flare*\\
0007&00 18 21.34&+30 57 20.6&1&1&0&0&0&0&G5&HD1405&9.6&8.6&RSCVn\\
0008&00 19 15.27&+07 48 20.7&1&1&0&0&0&0&M U&1RXS J001915.4+074816&0.0&0.0&            \\
\end{longtable}
\normalsize
\end{landscape}

\end{document}